\documentstyle[aps]{revtex}
\begin{document}
\twocolumn[
\preprint{DOE/ER/40762--051, UMPP \#95--058}
\title
{Baryon Mass Splittings in Chiral Perturbation Theory} %
\author{M. K. Banerjee and J. Milana}
\address
{Department of Physics, University of Maryland\\ College Park,
Maryland 20742, USA} %
\date{DOE/ER/40762--051, UMPP \#95--058, revised January 12, 1995.}
\maketitle

\begin{abstract}\widetext
Baryon masses are calculated in chiral perturbation theory at
the one--loop-${\cal O}(p^3)$ level in the chiral expansion
and to leading order in the heavy baryon expansion.  Ultraviolet
divergences occur requiring the introduction of counter--terms.
Despite this neccessity, no knowledge of the counter terms is
required to determine the violations to the Gell--Mann Okubo
mass relation for the baryon octet or to the decuplet equal
mass--spacing rule, as all divergences cancel {\it exactly} at
this order.  For the same reason all reference to an arbitrary
scale $\mu$ is absent.  Neither of these features
continue to higher--powers in the chiral expansion.
We also discuss critically the absolute neccessity of simultaneously
going beyond the leading order heavy baryon expansion, if one
goes beyond the one-loop-${\cal O}(p^3)$  level.
We point out that these corrections in $1/M_B$ generate new
divergences $\propto m^4/M_{10}$. These divergences together with
the divergences occuring in  one-loop-${\cal O}(p^4)$ graphs of chiral
perturbation theory are taken care of by the same set of counter--terms.
Because of these unknown  counter--terms  one cannot predict the
baryon mass splittings at the one-loop-${\cal O}(p^4)$ level.
We point out another serious problem of going to the one-loop-${\cal O}(p^4)$
level. When the decuplet is off its mass--shell there are additional
$\pi N\Delta$ and $\pi\Delta\Delta$ interaction terms. These interactions
contribute not only to the divergent terms $\propto (m^4/M_{10})$, but
also to nonanalytic terms such as $\propto (m^4/M_{10}){\rm ln}(m/M_{10})$.
Thus without a knowledge of the coupling constants appearing in these
interactions one cannot carry out a consistent
one-loop-${\cal O}(p^4)$ level calculation.
\end{abstract}

\vglue0.25in
]

\newpage
\narrowtext
\section{Introduction}
\label{Intro}
While chiral perturbation theory ($\chi$pt) has a long
history\cite{Pagels}, modern applications have been driven by
the formulation given by Weinberg in 1979.\cite{Wphilo} Using
power counting techniques, Weinberg demonstrated that for the
most general, non--linear chiral lagrangian in the purely
mesonic sector a loop expansion can be systematically developed
{\it even} though such lagrangians are not renormalizable in the
traditional sense.  Infinities generated by loops involving
terms of lower chiral power, a quantity which will be defined
shortly, are removed by terms of higher power in the lagrangian.
The systematics occur because higher power means higher order in
an expansion in terms of derivatives of the pion's field and the
pion's mass.  Provided one restricts kinematically the
application of the theory to scales of the order of the pion's
mass, $m$, such an expansion has at least the hope of converging. The
expansion parameter naturally occuring in this loop expansion is
$(m/2 \pi f_\pi)^2$.  Of course the introduction of
additional terms in the lagrangian requires additional
experimental information in order to fix the residual finite
piece of these higher power ``counter--terms''.  The number of
independent experimental inputs increases rather rapidly with
the loop--expansion.  For example, while the most general lowest
order chiral lagrangian in the mesonic sector, ${\cal L}_2$,
contains only two terms, there are ten independent terms at next
order, ${\cal L}_4$.  Nevertheless, nontrivial predictions
follow once these new terms are determined.  This program was
outlined by Weinberg in \cite{Wphilo}; its successful
implementation in the mesonic sector through the one loop level
was performed by Gasser and Leutwyler in their seminal papers of
the mid 1980s\cite{GassLeut}.

The extension of these methods to the nucleon sector was first
attempted by Gasser, Sainio and Svarc\cite{GassNuc}.  The
inclusion of baryons adds the nontrivial complication that the
nucleon mass $M$ is comparable to that of the typical chiral
scale $\chi \sim 2 \pi f_\pi$.  A loop expansion, when
calculated with the full nucleon propagator~\cite{GassNuc},
inevitably contains terms proportional to $M/\chi$ and powers
thereof.  Clearly one does not hope to form a convergent series
with such an expansion parameter.  Nevertheless the leading
infrared, ($m^2 \rightarrow 0$) nonanalytical behavior of
the graphs did appear in \cite{GassNuc} to be systematically
correlated with the loop expansion. The authors of
\cite{GassNuc} thus conjectured that this pattern would continue to all orders
in the loop expansion suggesting that such an expansion, if
organized properly, would be useful.

Weinberg ~\cite{Wcount} introduced the notion of chiral power.
A general $2N$ baryon legged graph is assigned the chiral power
$\nu$ given by the expression
\begin{equation}
\nu = 2 - N + 2L + \Sigma_i V_i (d_i + \frac{1}{2}n_i - 2),\label{powerc}
\end{equation}
in which $L$ is the number of loops, $V_i$ is the number of
vertices of type $i$ characterized by $d_i$ derivatives or
factors of $m$ and $n_i$ number of nucleon fields. The
systematic expansion required that the nucleon be considered
nonrelativistic.  To the extent that all relevant momentum are
of the order of the pion's mass, this constraint is consistent
with the entire program of chiral perturbation theory.
Weinberg's scheme validated the conjecture of
Ref.~\cite{GassNuc}.   We note that we use Eq. (\ref{powerc}) in all
further discussions to label the power of any particular graph.
Subsequent work of Weinberg\cite{WNNforce}
and others\cite{Kolck1} have focussed on the NN force.

By applying techniques developed for heavy quark
physics\cite{HeavyEff,pedagogy} to the baryon sector, Jenkins
and Manohar\cite{JenkMan1} formalized the nonrelativistic
treatment of the nucleon and made systematic counting of chiral
power possible.  All terms proportional to the nucleon's mass
are absent by construction, and the loop expansion in terms of
momentum and the pion's mass is realized.

The success of the chiral perturbation theory in the nucleon
sector relies on a double expansion: a chiral expansion in
$1/\chi$, and the heavy baryon expansion in $1/M_B$.  Among
graphs with the same number of $\pi N$ vertices these two
expansions are distinct in terms of the parameters of the QCD
lagrangian.  The chiral expansion is based on the mass of the
light quarks $m_u, m_d, m_s \rightarrow 0$, while the heavy
baryon expansion can be associated with the limit of large $N_c$
among these graphs.

The first comprehensive application of chiral perturbation
theory to the problem of octet and decuplet baryon masses is due
to Jenkins~\cite{Jenkins}. She examined the question why the
two well-known predictions, namely, the Gell-Mann Okubo~\cite{GMO}
(GMO) relation,
\begin{equation}
\frac34\,M_{\Lambda}+\frac14\,M_{\Sigma}-\frac12\,M_N-\frac12\,M_{\Xi}=0,
\label  {SU3GMO}
\end{equation}
and the Decuplet Equal Spacing Rule~\cite{nobel} (DES),
\begin{eqnarray}
(M_{\Sigma^*} - M_{\Delta}) - (M_{\Xi^*}-M_{\Sigma^*})& =&
\nonumber \\
(M_{\Xi^*} - M_{\Sigma^*}) - (M_{\Omega^-}-M_{\Xi^*})& =&
\nonumber \\
\frac12\{(M_{\Sigma^*} - M_{\Delta}) -(M_{\Omega^-}-M_{\Xi^*})\} &&
\label {SU3DES}
\end{eqnarray}
work as well as they do despite
apparently large corrections coming from the one-loop-${\cal O}(p^3)$
level. The experimental value of the left-hand side of
Eq.~(\ref{SU3GMO}) is $6.5\,MeV$ which is $~3\%$ of the average
intra-multiplet splitting among the octets. The average
experimental value of the mass combinations in
Eq.~(\ref{SU3DES}) is $27\,MeV$ which is $~20\%$ of the average
intra-multiplet splitting among the decuplets.  We remind the
reader that the two predictions above are based on the assumption
that the flavor symmetry breaking term in the lagrangian
transforms like the $\lambda_8$ member of an octet (which is true for QCD)
and that its effect may be derived perturbatively.  Jenkins went
up to ${\cal O}(p^4)$ level by
inserting octet and decuplet sigma terms in the loop diagrams
and stressed the importance of these terms in explaining the
susprising success of GMO and DES.

In this paper we reexamine the application of chiral
perturbation theory to the problem of octet and decuplet baryon
masses.  We use the heavy baryon formalism of Jenkins and Manohar
to include the decuplet field\cite{JenkMan2}, and also many of the
useful tables which appear  in Ref.~\cite{Jenkins}.
We differ from Jenkins on two points,  one major and one minor.
We also report a new result concerning $1/M_B$ corrections to the
heavy fermion theory.  The three points  are listed below.
\begin{enumerate}
\item As we will see later, divergences occur at the one-loop-${\cal O}(p^3)$
level when the internal baryon and the external baryon are in different
flavor multiplets. The resulting counter--terms are
combinations of flavor singlet and flavor octet
(specifically, the $\lambda_8$ member).
Nevertheless one can predict the results of
GMO and DES, because the flavor structure of the
counter-terms ensures that they do not contribute to these mass combinations.
We also note that the counter--terms have structures similar to those appearing
in ${\cal L}^0$ and ${\cal L}^1$, but have higher chiral power,
namely, ${\cal O}(p^2)$.  When one goes to the one-loop-${\cal O}(p^4)$ level
by
inserting octet and decuplet sigma terms one
needs two types of counter--terms not present in
${\cal L}^0$ or ${\cal L}^1$. The wavefunction renormalization
counter--terms arising from Fig.~1a are ${\cal O}(p^2)$ flavor
octets. They contribute through diagrams containing one of these terms
and a sigma term separated by a baryon propagator.  The net
effect belongs to flavor
$8\otimes8=1\oplus 8\oplus 8\oplus 10\oplus\bar{10}\oplus 27$ space and
contributes to GMO and DES. The vertex renormalization graphs shown in Fig.~1b
 generate counter-terms of ${\cal O}(p^3)$ proportional
to the square of the quark mass matrix. Hence they also  belong to flavor
$1\oplus 8\oplus 8\oplus 10\oplus\bar{10}\oplus 27 $ and will also contribute
to GMO and DES. The
most important point is that these counter-terms are not of
forms already present in ${\cal L}^0$ or ${\cal L}^1$.
Unlike multiplicatively renormalizable theories where one can
discuss various regularization schemes
(e.g. MS or $\overline{{\rm MS}}$) without changing
the underlying number of inherent parameters in the theory, these
counter--terms have residual finite pieces that require further experimental
input to determine.
We do not know the relevant coupling
constants and, hence, cannot predict the values of the GMO or
DES mass combinations. We need experimental values of these
combinations and other experimental data to determine the unknown
coupling constants. Thus we are forced to conclude that so far
as GMO and DES are concerned we have at present power to predict
only up to the one-loop-${\cal O}(p^3)$ level and not beyond that.

We calculate the left hand sides of
Eqs.~(\ref{SU3GMO}) and (\ref{SU3DES}) at  the one-loop-${\cal
O}(p^3)$ level only.  In principle, these results are thus contained, at
least in part (see point 2 below) in the results
of Ref.~\cite{Jenkins}, but not explicitly identified. It is
important to know what the results are at one-loop-${\cal
O}(p^3)$ level because we find that this is the limit of
predictability of chiral perturbation theory in the area of
baryon masses.

We note that the GMO results at the one-loop-${\cal O}(p^3)$
level have been published already by Bernard {\it et al}~\cite{BerMeis1}.
Similar results for the DES, contained here, are new.

\item The physical value for the mass difference between the decuplet
and octet baryons, $ M_{10} - M_{8} \approx 2 m$,
and it should share with $m$ the chiral power $1$.  Then,
according to Eq.~(\ref{powerc}), the decuplet-octet mass
difference term in the lagrangian has chiral power $0$ and is
included in $L^0_v$ of Jenkins~\cite{Jenkins}.   The value of a
baryon-meson loop is expressed with the help of a function
$W(m,\delta,\mu)$ parametrized by the meson mass,  $m$,
the renormalization scale,  $\mu$, and the quantity $\delta$
defined below:
\begin{eqnarray}
{\rm Octet-octet}&&\,\,\,\,\,\,\delta=0, \nonumber \\
{\rm Octet-decuplet}&&\,\,\,\,\,\,\delta=M_{10}-M_8,  \nonumber \\
{\rm Decuplet-octet}&&\,\,\,\,\,\,\delta=M_8-M_{10},  \nonumber \\
{\rm Decuplet-decuplet}&&\,\,\,\,\,\,\delta=0. \label{delta}
\end{eqnarray}
The first label on the left of each line is for the external leg
and the next one is for the internal leg.
This function, defined by Eqs.~(\ref{del0}), (\ref{mgtdel}) and (\ref{delgtm}),
has a branch point at $\delta= \mp m$, reflecting the instability of the
decuplet
(octet) to decay into an octet (decuplet) and a meson when the masses allow the
process.  Because of the proximity of the
branch point and because both GMO and DES involve cancellation among
large quantities, we argue that the role of $\delta$ should not be
treated perturbatively~\cite{Jenkins}.
\footnote{There are circumstances where such a treatment is appropriate,
 as in the case of isospin splittings discussed recently by Lebed\cite{Lebed}.}
It should be included in all
orders~\cite{gangof4}.    In Table~1, which appears later in the paper, where
we justify our argument, we show that the difference between the results of
perturbative and exact treatments of the $\delta$ term is of the order of the
experimental values of   the left hand sides of
Eqs.~(\ref{SU3GMO}) and (\ref{SU3DES}).

We note that interesting questions concerning the two limits,
$\delta \rightarrow 0$ and $m \rightarrow 0$,
in the context of large $N_c$ have been discussed by Cohen and
Broniowski\cite{XptandNc}.

\item The leading $1/M_B$ corrections to the heavy fermion theory
results is $\sim (m^4/M_B)$. We find that when the internal baryon
is a decuplet, the $1/M_B$ corrections to the one-loop result is
actually divergent.
Specifically, it has the form
$\sim (m^4/M_{10})(\frac{1}{\epsilon}-\gamma_E+{\rm ln}\,(4\pi))$.
This result has important implications. It means that one must add
counter-terms
$\sim (m^4/M_{10})$ to be fixed with the help of experimental data. It is not
possible to calculate the $1/M_B$ correction terms {\it ab initio}. We have
seen
earlier that  counter-terms  $\sim m^4$  are needed when one goes to  the
one-loop-${\cal O}(p^4)$ level by inserting   sigma terms into
one-loop-${\cal O}(p^3)$ graphs.   Having the same flavor $SU(3)$ group
structure both counter-terms will be determined together from the same
experimental information, namely, the octet and decuplet masses. We cannot
separate the contributions to the experimentally fixed counter-term from the
two mechanisms - $1/M_B$ corrections and sigma term insertions.

We discover an additional complication for future one-loop-${\cal O}(p^4)$
level calculations. The $\pi N\Delta$\cite{Nath1} and $\pi\Delta\Delta$
couplings each contain an additional term which is $1/M_B$ suppressed compared
 to the term retained in the heavy fermion theory.  These terms
contribute to the ultraviolet divergent term in $\frac{m^4}{M_{10}}$.
This by is not a matter of concern. As we have noted above one can only
fix the strength of the total $\sim m^4$ counter-term and not the part coming
from $1/M_B$ effects. But these divergences are also accompanied by finite,
nonanalytic terms in $m/M_B$.  Such terms must thus be calculated and
included in the
expressions for the baryon masses used to  determine the counter-terms.
But it cannot be done without knowing the values of the secondary coupling
constants.  As these coupling constants play their roles only when the $\Delta$
is
off its mass shell,  to fix them reliably from experiment in a credible manner
may prove to be a nearly impossible task.  The point is illustrated by the work
of
Benmerrouche,  Davidson, and   Mukhopadhyay~\cite{nimai}.
They   attempted to fix  the secondary coupling constant $\alpha$,
defined later in Eq.~(\ref{theta}),  which appear in $\pi N\Delta$ interaction
and were able  only  to place its value within a rather broad range,
namely, $0.30\geq\alpha\geq -0.78$.
 \end{enumerate}

The lowest order lagrangian depends upon four coupling
constants: $D$ and $F$ describe meson--octet couplings, ${\cal
C}$ baryon octet--decuplet couplings and, ${\cal H}$
meson--decuplet couplings.  The values of the first three are
reasonablely well determined.  The GMO combination of masses,
depends only on these quantities and, using the values of
Jenkins~\cite{Jenkins}, we obtain typically $9\,MeV$ while the
experimental value is $6.5\,MeV$.  The value of the decuplet
spacing depends upon ${\cal H}$ which is difficult to determine
experimentally.  If we chose to fit the average violation to the
decuplet equal spacing rule ($27\,MeV$) at the one-loop-${\cal O}(p^3)$  level
we obtain ${\cal H}^2 = 6.6$.

\section{Baryon Self--Energies}
\label{BSE}
\subsection{{\it The purely Octet sector}}
\label{Octet}
Up to  ${\cal O}(p^3)$ the effective chiral lagrangian coupling
octet pseudoscalar mesons to octet baryons
is:\cite{pedagogy,Jenkins}
\begin{eqnarray}
{\cal L}_{eff} &=& {\cal L}_0^{\pi N} + {\cal L}_1^{\pi N} +
{\cal L}_2^{\pi \pi}
\nonumber\\
{\cal L}_0^{\pi N} &=& Tr \overline{B} (i\not\!\!D - M_B) B +
\nonumber\\
&&D Tr \overline{B} \gamma^\mu \gamma_5 \{A_\mu, B\} + F Tr
\overline{B} \gamma^\mu \gamma_5 [A_\mu,B] \nonumber\\
{\cal L}_1^{\pi N} &=& b_D Tr \overline{B} \{\xi^\dagger M
\xi^\dagger + \xi M \xi, B\}
\nonumber\\
&&+ b_F Tr \overline{B} [\xi^\dagger M \xi^\dagger + \xi M \xi,B]\nonumber\\
&&+ \sigma Tr M(\Sigma + \Sigma^\dagger) Tr \overline{B} B
\nonumber\\
{\cal L}_2^{\pi \pi} &=& \frac{f_\pi^2}{4} Tr \partial_\mu \Sigma
\partial^\mu \Sigma^\dagger
+ a Tr M(\Sigma + \Sigma^\dagger),
\label{lagfull}
\end{eqnarray}

\noindent in which,
\begin{eqnarray}
&\xi = e^{i \pi/f_\pi},
\hspace{.5in}
&\Sigma = \xi^2 = e^{i 2\pi/f_\pi},\nonumber\\
&V_\mu = \frac{1}{2}
[(\partial_\mu\xi) ^\dagger\xi + \xi^\dagger
(\partial_\mu \xi)],
&A_\mu = \frac{i}{2}[(\partial_\mu\xi) ^\dagger\xi - \xi^\dagger
(\partial_\mu \xi)]  ,\nonumber\\
&D^\mu B = \partial^\mu B + [V^\mu, B].\hspace{.3in}&
\end{eqnarray}
The definitions of the mass matrix, $M$, and the octet meson and
baryon fields are, by now, standard, and are given
in Ref.~\cite{pedagogy,Jenkins}.  Note that the subscripts on the
baryonic sector of ${\cal L}_{eff}$ refer to the chiral power
defined by Eq.~(\ref{powerc}).

The one loop nucleon self--energy, $\Sigma(p,M_B)$, is shown
diagramatically in Fig. (2).  The expression for $\Sigma(P,
M_B)$ on mass--shell is given by
\begin{eqnarray}
&&\Sigma(P, M_B) = \frac{i\beta}{2 f_\pi^2}\int \frac{d^4
k}{(2\pi)^4}\frac{\gamma_5 \not\!k (\not\!P + \not\!k + M_B)
\gamma_5 \not\!k}{(k^2 - m^2_\pi + i\eta)(2P \cdot k + k^2 + i\eta)}
\nonumber\\
&&= \frac{-i\beta}{2 f_\pi^2}\int \frac{d^4 k}{(2\pi)^4}
\frac{ (M_B + \not\!P) k^2}
{(k^2 - m^2_\pi + i\eta)(2P \cdot k + k^2 +
i\eta)},\label{selfE1}
\end{eqnarray}
where $\beta$ represents SU(3) algebra factors.

The heavy baryon result\cite{Jenkins} for $\Sigma(P,M_B)$ can be
obtained by introducing $P = m v$ and taking the $M_B
\rightarrow \infty$ limit of the
{\it integrand} in the above, whereby one obtains that
\begin{equation}
\Sigma(P, M_B \rightarrow \infty) =
\frac{-i\beta}{2 f_\pi^2}\int \frac{d^4 k}{(2 \pi)^4} \frac{\frac12(1 +
\not\!v) k^2}
{(k^2 - m^2_\pi + i\eta) (v \cdot k + i\eta)}.\label{mshift1}
\end{equation}
The same result is obtained by first reducing the effective
lagrangian ${\cal L}_{eff}$ in the heavy fermion limit in terms
of velocity fields $B_v$,\cite{pedagogy}
\begin{equation}
\frac{1}{2}(1 + \not\!v) B(x) = e^{-iM_B v\cdot x} B_v(x).\label{wvf}
\end{equation}
Any reference in ${\cal L}_{eff}$ to $M_B$ is thereby removed, so
that, for example, ${\cal L}_0^{\pi N}$ becomes\cite{JenkMan1}
\begin{eqnarray}
{\cal L}_{v}^0 &=& i Tr \overline{B}_v v\cdot D B_v +
\nonumber\\
&&2 D Tr \overline{B_v} S^\mu_v \{A_\mu, B_v\} +2 F Tr
\overline{B_v} S^\mu_v [A_\mu, B_v]. \label{Lheavy}
\end{eqnarray}
where $S^\mu_v$ is a spin factor defined in
Refs.\cite{JenkMan1,JenkMan2}.  Observe that from ${\cal
L}^0_v$, the nucleon's propagator is given directly to be
$i/(v\cdot k + i\eta)$.

As in previous works,\cite{GassNuc,Jenkins} we use dimensional
regularization to evaluate all integrals.  In the purely mesonic
sector it is well known that dimensional regularization, by not
introducing any additional mass parameters, avoids
complications\cite{Gerstein} in the path--integral arising from
the chiral--invariance of the measure.  We know of no such
similar result involving the baryon sector but find that the use
of alternative regularization schemes such as Euclidean cutoff,
that introduces additional mass parameters, would complicate the
power counting result of Weinberg, Eq.  (\ref{powerc}).  In
order to avoid these complications we use dimensional
regularization.

In  Appendix A we present one method of evaluating Eq.
(\ref{mshift1}).  One finds that the mass--splitting $\delta
M_B$ is given simply in the heavy baryon limit by
\begin{equation}
\delta M_B\left |_{M_B \rightarrow \infty} = \beta \frac{- m^3}{16
\pi f^2_\pi}\right. . \label{delmb}
\end{equation}

The wavefunction normalization, $Z_2$,  is given by the following expression.
\begin{equation}
Z_2^{-1} =1 - \frac{ m^2}{8 \pi^2f^2_\pi} \left(
\frac{1}{\epsilon} -
\gamma_E + 1 + {\rm ln}(4\pi) -
{\rm ln}\frac{m^2}{\mu^2} \right).
\end{equation}
Clearly $Z_2$ requires renormalization which is accomplished
through counterterms of chiral power 2 in ${\cal L}_{eff}$.  These
have been given by Lebed and Luty~\cite{LebedLuty}.   These authors have
suggested  that the wavefunction renormalization
counterterms   may be absorbed by
redefining the baryon field.  Although this is correct for the
$Tr\,\bar{B}iD\!\!\!\!/B$ term in the lagrangian, such a redefinition will
neccessarily generate new interaction terms.  For example,
otherwise charge conservation, which requires cancellation
between wavefunction renormalization and vertex renormalization
($Z_1=Z_2$), cannot be maintained.  Thus the need to deal with
these counterterms cannot be avoided.
Use of only the logarithmic piece in $Z_2$ is
not sufficient~\cite{JenkMan1,Jenkins,JenkMan2,gangof4,ButSav}.

For the present case,  by confining ourselves to only
the one-loop-${\cal O}(p^3)$ level,  we    avoid the
complications of the wavefunction
renormalization  as well as the $1/M_B$
corrections discussed earlier.

We will now discuss the inclusion of the decuplet, which
involves its own unique features.

\subsection{{\it The Decuplet}}
The decuplet is included as a spin $3/2$ Rarita--Schwinger
field\cite{RaritaS} $\Delta^\mu$.  On--shell, $\Delta^\mu$ obeys
the Dirac equation
\begin{equation}
(i \not\!\partial - M_{10}) \Delta^\mu = 0
\end{equation}
along with the constraints
\begin{eqnarray}
\gamma_\mu \Delta^\mu &=& 0,\nonumber\\
\partial_\mu \Delta^\mu &=& 0, \label{constraints}
\end{eqnarray}
which eliminate the spin $1/2$ components of the $\Delta^\mu$
field.  The most general free lagrangian for $\Delta^\mu$ that
generates the Dirac equation of motion and the constraints
is\cite{Nath1,MoldC,nimai,Griegel}
\begin{eqnarray}
{\cal L}_{\Delta} &= -\overline{\Delta^\mu}[ (i \not\!\partial -
&M_{10}) g_{\mu \nu} + i A(\gamma_\mu \partial_\nu + \gamma_\nu
\partial_\mu)\nonumber\noindent\\
&&+\frac{1}{2}(3A^2 + 2A + 1) \gamma_\mu \partial^\alpha
\gamma_\alpha \gamma_\nu\nonumber\\
&&+M_{10}(3A^2 + 3A + 1)\gamma_\mu \gamma_\nu ] \Delta^\nu,
\label{deltafree}
\end{eqnarray}
where $A$ is an arbitrary (real) parameter subject to the one
requirement that $A \ne -1/2$.  Taking $A = -1$ leads to the
most commonly used expression for the decuplet propagator,
\begin{eqnarray}
G^{\mu \nu} &= \left. \frac{1}{i}\frac{\not\!P + M_{10}}{P^2 - M_{10}^2 +
i\eta}
\right[ g^{\mu \nu} - \frac{1}{3} \gamma^\mu \gamma^\nu
&-\frac{2}{3} \frac{P^\mu P^\nu}{M_{10}^2}\nonumber\\ &&\left. +
\frac{P^\mu \gamma^\nu - P^\nu \gamma^\mu}{3 M_{10}} \right]. \label{propdec}
\end{eqnarray}
To leading order in the heavy baryon expansion,
where\cite{HeavyEff,JenkMan2} one takes $P = M_{8} v + k$, the
decuplet propagator becomes
\begin{eqnarray}
G_v^{\mu \nu} &= \left. \frac{1}{i}\frac{\frac{1}{2}(1 + \not{v})} {v\cdot
k - \delta + i\eta}
\right[ g^{\mu \nu} - \frac{1}{3} \gamma^\mu \gamma^\nu &- \frac{2}{3}
v^\mu v^\nu\nonumber\\ &&\left. + \frac{1}{3}(v^\mu \gamma^\nu -
v^\nu\gamma^\mu)
\right]\nonumber\\
&\equiv \frac{1}{i}\frac{\frac{1}{2}(1 + \not{v})}{v\cdot k - \delta +
i\eta}P_v^{\mu \nu}.&
\label{decprop}
\end{eqnarray}
The quantity $\delta$, which we take to be $226\,MeV$, is the mass
difference between the baryon octet and baryon decuplet masses.

The constraints on the decuplet field in the heavy baryon theory have
been given by Jenkins and Manohar~\cite{JenkMan2}
\begin{eqnarray}
\gamma_\mu\Delta^\mu&=&0, \nonumber\\
v_\mu\Delta^\mu&=&0. \label{hfconst}
\end{eqnarray}

The most general, chirally invariant interaction lagrangian
involving decuplets, octet baryons and octet mesons is:
\begin{eqnarray}
{\cal L}^i &= {\cal C} (\overline{\Delta^\mu} \Theta_{\mu \nu}
A^\nu B +& h.c.) +{\cal H} \overline{\Delta^\mu} \gamma_\nu
\gamma_5 A^\nu \Delta_\mu\nonumber\\
&&+ \tilde{H} (\overline{\Delta^\mu} \gamma_\mu \gamma_5 A^\nu
\Delta_\nu + h.c.) \label{Lint}
\end{eqnarray}
where $\Theta_{\mu \nu}$ is given by\cite{Nath1,nimai}
\begin{equation}
\Theta_{\mu \nu} = g_{\mu \nu} + \alpha \gamma_\mu \gamma_\nu. \label{theta}
\end{equation} In the heavy fermion theory the last term vanishes
and the first two terms beome the interaction terms~\cite{JenkMan2}:
\begin{equation}
{\cal L}_v^i = {\cal C} (\overline{\Delta^\mu} A^\mu B + h.c.) +
2{\cal H} \overline{\Delta^\mu} S_{v\nu} A^\nu \Delta_\mu. \label {LI10v}
\end{equation}

Using the decuplet propagator given by
Eq. (\ref{decprop}) and the constraints given by Eq.~(\ref{hfconst}),
one obtains for the mass--shift from Fig. (3)
\begin{eqnarray}
\delta M^\prime &=& \frac{-i3 \beta^\prime}{4 f_\pi^2}
\int \frac{d^4 k}{(2 \pi)^4} \frac{k_\mu k_\nu P_v^{\mu \nu}}
{(k^2 - m^2  + i\eta)(v \cdot k - \delta + i\eta)}, \nonumber \\
&=& \frac{\beta^\prime}{16 \pi f^2_\pi} \left[
\frac{-3 \delta}{2 \pi}
( m^2  - \frac23\delta^2 ) ( \frac{1}{\epsilon} - \gamma_E +
{\rm ln}(4\pi) - {\rm ln}\frac{m^2 }{\mu^2} ) \right. \nonumber\\
&-&\left. \frac{3 \delta m^2}{2 \pi} - \frac{2}{\pi}(m^2 -
\delta^2)^{3/2}
{\rm tan}^{-1}\frac{\sqrt{m^2-\delta^2}}{\delta} \right]
.\label{uvterms}
\end{eqnarray}

It is clear from Eq. (\ref{uvterms}), and as announced earlier, that upon
inclusion of the decuplet, the mass--shift requires renormalization.  Two
types of counter-terms belonging to ${\cal L}_2^{\pi N}$ are needed,
one to cancel the divergence proportional to $\delta^3$ and the
other in $\delta m^2 $.  The $\delta^3$ term turns into an
overall mass--shift when all {\it relevant} intermediate states
are summed over.  The $\delta m^2 $ term is a sum of flavor singlet and
flavor octet.  As noted earlier all counterterms (divergences) cancel
{\it exactly} in the mass combinations which appear in the
GMO and the DES. For completeness,  the counter-terms are listed in Appendix B.

\section{$1/M_B$ Corrections}
If in subsection~\ref{Octet}, instead of adopting the heavy fermion theory,
we had evaluated Eq.~(\ref{selfE1}) we would have obtained the
following expression for the octet self-energy contribution coming from a
loop containing an octet baryon internal line:

\begin{eqnarray}
\delta M_B &=& \frac{\beta}{16 \pi f^2_\pi}
\left[\frac{M_B^3}{\pi}\left(\frac{1}{\epsilon} - \gamma_E + {\rm ln}\,(4\pi) +
1
- {\rm ln}M_B^2\right)\right.\nonumber\\ &&+\frac{M_B m^2}{
\pi}\left(\frac{1}{\epsilon}
- \gamma_E + {\rm ln}\,(4\pi) + 2 - {\rm ln}M_B^2\right)\nonumber\\
&&\left.-m^3\left(1 - \frac{m}{\pi M_B}\left[1 + {\rm ln}
\frac{M_B}{m}
\right]\right)+\cdots\right].
\label{fullmass}
\end{eqnarray}
The ultraviolet divergences proportional to $M^3_B$ and $M_B$
are those first noted by Gasser, Sainio and Svarc\cite{GassNuc}.
The additional divergent terms obtained by letting $M_B\rightarrow\infty$
have identical flavor structure.  Observe that no nonanalytical behavior
in $m_\pi$ is thus lost in the $M_B\rightarrow\infty$ limit.

The new information contained in Eq.~(\ref{fullmass}) is the $1/M_B$
correction term in the last line. The contribution of the correction term to
GMO is $\sim40\%$ of that of the $m^3$ term. Thus it is quite substantial.
Unfortunately, it is not enough to include finite, chirally nonanalytic terms
like these to obtain the leading $1/M_B$ corrections.
The reason is that a one-loop graph containing a decuplet internal line gives a
divergent contribution proportional to $m^4/M_B$. They arise from the presence
of
the $-\frac23\frac{P^\mu P^\nu}{M^2_{10}}+
\frac{P^\mu \gamma^\nu - P^\nu \gamma^\mu}{3 M_{10}} $
terms in the decuplet propagator given by Eq.~(\ref{propdec}).
Within a loop $P=p+k$, where $p$ is the external momentum and
$k$ the loop momentum. The appearance of   extra powers of the loop
momentum is responsible for the divergent $1/M_B$ terms.

The interaction terms
${\cal C} \alpha (\overline{\Delta^\mu} \gamma_\mu\gamma_\nu A^\nu B
+ h.c.)$ and
$\tilde{H} (\overline{\Delta^\mu} \gamma_\mu \gamma_5 A^\nu \Delta_\nu +
h.c.)$,
  which appear in Eq.~(\ref{Lint}), also contribute to the divergent terms.
The combined results are shown below.
$$\delta M_8 : \frac{-1}{64 f_\pi^2 \pi^2}
[\frac{(M_8+M_{10})}{M_{10}}(1+\alpha)(2-\alpha)-3 \alpha(1+\alpha)]$$
$$\hspace{1in} \sum_\lambda \,
\beta^\prime_8(\lambda)\frac{m^4_\lambda}{M_{10}}
[\frac{1}{\epsilon}-\gamma_E + {\rm ln} (4\pi)]$$
$$\delta M_{10} : -\frac{3 [1 - 2 \tilde{H}/{\cal H}  - 3 (\tilde{H}/{\cal
H})^2]}
{320 f_\pi^2 \pi^2}\hspace{1.5in}$$
\begin{equation}
\hspace{1in} \sum_\lambda \, \beta_{10}(\lambda)\frac{m^4_\lambda}{M_{10}}
[\frac{1}{\epsilon}-\gamma_E + {\rm ln} (4\pi)]
\end{equation}

The presence of these counter-terms has four important consequences.

First, we can no longer calculate the $1/M_B$ corrections completely.

Second, the flavor structure of the divergences and, hence, of the
counter-terms
is identical to those appearing at the one-loop-${\cal O}(p^4)$ level due to
the
insertion of   sigma terms into one-loop-${\cal O}(p^3)$ graphs.
The sum of the two groups of counter-terms have to be fixed with the help of
experimental data, which may include the octet and decuplet mass
 splittings themselves.

Third, since the two share the same counter--term,
when one goes to the one-loop-${\cal O}(p^4)$ level in chiral perturbation
theory one must also include the leading $1/M_B$ corrections.
The finite, nonanalytic terms from both sources must be
regarded, {\it a priori}, as equally important.  This leads us to the
next consequence.

The fourth consequence may prove to be a serious impediment to any
one-loop-${\cal O}(p^4)$ calculation.  The $m^4/M_{10}$ divergent terms are
inevitably accompanied by nonanalytic terms,
e.g. $(m/M_{10})^4 {\rm ln}(m/M_{10})$ in
Eq.~(\ref{fullmass})).  These terms must be calculated
and included in expressions used to fix the full $m^4$ counter-terms.
Unfortunately, these log terms depend on the quantities $\alpha$ and
$\tilde{H}$.
As stated earlier, these coupling constants affect physical results
only through the
role of virtual $\Delta$s.  Until these constants can be reliably determined
from
experiment, we are at an impasse in the application of chiral
perturbation theory in the nucleon sector.

Anatomy of the $1/M_B$ divergences suggests that the pattern will continue to
chiral perturbation theory results involving more loops and powers. The $1/M_B$
correction to the heavy fermion theory result at any level will involve
divergent
terms of a level with one more
chiral powers, coming always from internal decuplet lines.
These divergences will always be inseparably entangled with chiral
divergences of the
same level in heavy fermion theory. Thus the $1/M_B$ correction terms cannot
be calculated without a complete calculation at the same level of chiral power.
It
must be noted that this requirement is imposed not merely by considerations of
consistency but by the appearance of divergences.
One is not left with any option in the matter.

\section{Mass Splittings}
Most of the algebraic quantities used in this section have appeared in
Refs.~\cite{Jenkins,gangof4,JenkManconf}.   For the reader's convenince we
include the coefficients $\alpha_i$ and $\beta_i$ defined by
Jenkins~\cite{Jenkins}.
They appear in Tables~\ref{tablealp} and ~\ref{tablebet}.

Following the style of Ref. \cite{Jenkins} we write for the
mass, $M_i$, of the ``ith'' baryon through the one--loop order as
\begin{eqnarray}
M_i &=& M_B +\frac12(1 \mp 1)\delta + \alpha_i +
\alpha_i^\delta (\mu)
\frac{3 \delta\zeta}{32 \pi^2 f_\pi^2} \nonumber\\ &&-\Sigma_{\lambda
} \beta_i(\lambda) \frac{m_\lambda^3}{16 \pi f^2_\pi}
- \frac{\delta^3}{16 \pi^2 f^2_\pi} a_R^{\pm, \Delta}(\mu) \nonumber\\
&&- \Sigma_{\lambda} \beta^\prime_i(\lambda)
W(m_\lambda,\delta,\mu). \label{massformula}
\end{eqnarray}
In the second term the upper (lower) sign is for octet (decuplet) baryons.
The coefficients $\alpha_i$ come from ${\cal L}_1^{\pi
N}$.  The term proportional to $\beta_i$ arises from the chiral
loops in Fig. 2 in which the propagating baryon is in the same
multiplet as as the baryon $i$, while the terms proportional to
$\beta^\prime_i$ arises from the loops in Fig. 3 where the
propagating baryon comes from the other multiplet.  The
sum over $\lambda$ runs over $\pi$, $K$ and $\eta$ mesons.   The
quantities $\alpha_i^\delta$, are obtained by adding a
superscipt $\delta$ to each of the entries in
Table~\ref{tablealp}.  The set $\{b_D^\delta, b_F^\delta, \cdots\}$,
thus generated,  is defined in the appendix.  $\zeta$ is the proportionality
constant in the GMOR\cite{GMOR} relation,
$\zeta = m^2_K/(m_s + \tilde{m}) = m^2_\pi/2\tilde{m}$.

The coefficents $a_R^{N, \Delta}(\mu)$ and
$\alpha_i^\delta (\mu)$ depend implicitly upon a
choice of scale, $\mu$.  This scale appears explicitly in the function
$W(m,\delta,\mu)$\cite{BerMeis1,JenkManconf}.  We
give below the expressions for the function for three cases of
interest:
\begin{eqnarray}
\delta=0,\,\,\,\, && W(m,\delta, \mu)
=\frac{1}{16 \pi f^2_\pi}m^3, \label{del0}\\
m>\mid\delta\mid,\,\,\,\, && W(m,\delta,\mu)
=\nonumber\\ &&\frac{1}{8 \pi^2 f^2_\pi}(m^2 - \delta^2)^{3/2}
{\rm tan}^{-1}\frac{ \sqrt{m^2-\delta^2}}{\delta} \nonumber\\
&&- \frac{3\delta}{32 \pi^2
f^2_\pi}\left(m^2-\frac{2}{3}\delta^2\right)
{\rm ln}\frac{m^2}{4\delta^2}\nonumber\\ &&-\frac{3\delta}{32
\pi^2
f^2_\pi}\left(m^2-\frac{2}{3}\delta^2\right)\,{\rm ln}\frac{4\delta^2}{\mu
^2},\label{mgtdel} \\
\mid\delta\mid>m,\,\,\,\, &&W(m,\delta,\mu) =\nonumber\\
&&\frac{-1}{16 \pi^2f^2_\pi}(\delta^2-m^2)^{3/2}
{\rm ln}\frac{\delta-\sqrt{\delta^2-m^2}}{
\delta+\sqrt{\delta^2-m^2}}\nonumber\\
&&- \frac{3\delta}{32 \pi^2
f^2_\pi}\left(m^2-\frac{2}{3}\delta^2\right)
{\rm ln}\frac{m^2}{4\delta^2}\nonumber\\ &&-\frac{3 \delta}{32
\pi^2
f^2_\pi}\left(m^2-\frac{2}{3}\delta^2\right)\,{\rm ln}\frac{4\delta^2}{\mu
^2},\label{delgtm}
\end{eqnarray}

When the SU(3) algebra factors are   included
the $W$'s appear in the combination
\begin{eqnarray}
V(\delta)&=&\left\{-\frac14\,W(m_\pi,\delta, \mu)+\right.\nonumber\\
&&\left.W(m_K,\delta,\mu)-
\frac34\,W(m_\eta,\delta,\mu)\right\} \label{Vcomb}
\end{eqnarray}

In the present order of the chiral expansion the octet meson
mass squares are taken to be proportional to the masses of the
current quarks~\cite{GMOR}:
\begin{equation}
m^2_\eta = \frac{4}{3}m^2_K-\frac13\,m^2_\pi.
\label{etamass}
\end{equation}
We note from the Eqs.~(\ref{del0}), (\ref{mgtdel}) and
(\ref{delgtm}) that the ${\rm ln}\,\mu^2$ term appears in
these expression with factors which are linear in $m^2$.
Combining this fact with Eq.~(\ref{etamass}) it is easy to see
that $V( \delta)$ does not contain terms proportional to
${\rm ln}\,\mu^2$.

{}From Eq.~(\ref{delgtm}) one finds that when $m\rightarrow\,0$
the quantity
\begin{eqnarray}
W(m,\delta,\mu)&&\simeq\,\frac{-1}{16\pi^2f^2_\pi}
\left[\frac32\delta(m^2-\frac23\delta^2)\,{\rm
ln}\frac{4\delta^2}{\mu^2}\right.
\nonumber\\
&&\left.
+\frac12\delta\,m^2-\frac{9}{16}\frac{m^4}{\delta}+\frac38\,\frac{m^4}
{\delta}\,{\rm ln}\frac{m^2}{4\delta^2}\right].
\end{eqnarray}
Thus it is perfectly well-behaved.

Finally one obtains for the Gell--Mann Okubo mass relation the
expression
\begin{eqnarray}
\frac34\,M_\Lambda + &&\frac14\,M_\Sigma-\frac12\,M_N-\frac12\,M_\Xi
=\nonumber\\ &&\frac23 \, (D^2-3F^2)V(0)-\frac19\,{\cal C}^2
V(\delta),\label{GMOf}
\end{eqnarray}
and for the violations to the Decuplet Equal Spacing rule:
\begin{eqnarray}
&(M_{\Sigma^*} -& M_{\Delta}) - (M_{\Xi^*}-M_{\Sigma^*})
=\nonumber\\
&(M_{\Xi^*} -& M_{\Sigma^*}) - (M_{\Omega^-}-M_{\Xi^*})
=\nonumber\\
&\frac12\{(M_{\Sigma^*} -& M_{\Delta}) -(M_{\Omega^-}-M_{\Xi^*})\}
=\nonumber \\
&&\frac29{\cal C}^2 V(\delta)-\frac{20}{81}\, {\cal H}^2 V(0).
\label{DESf}
\end{eqnarray}
We remind the reader of our convention, Eq. (\ref{delta}), by which
$\delta$ is a negative quantity in Eq. (\ref{DESf}).
We also remind the reader that all counterterms have explicitly
cancelled in these two relations and that the relations are
independent of the scale $\mu$.

Before discussing the numerical results following from
Eqs.~(\ref{DESf}) and (\ref{GMOf}) and their implications, we
comment on the accuracy of a perturbative evaluation of the
combination $V(\delta)$, Eq. (\ref{Vcomb}).  The perturbative
value is obtained by expanding the combination in a power series
of $\delta$ and retaining only the terms independent of $\delta$
and linear in $\delta$.  In Table~\ref{pert} we compare GMO and
DES using the exact and the perturbative values of
$V(\delta)$ using the parameter set of Ref. \cite{Jenkins}
given in set 1 of Table~\ref{tablex} below.
\begin{table}
\begin{center}
\begin{tabular}{|c|c|c|}
Quantites&Exact&Perturbative\\ \hline $GMO(MeV)$&$10.0$&$-1.1$\\
\hline
$DES(MeV)$&$-4.2$&$24.2$\\
\end{tabular}
\end{center}
\caption{$GMO$ and $DES$ using exact and perturbative values of $V(\delta)$.}
\label{pert}
\end{table}
It is clear that the differences are comparable to the
experimental values of the mass combinations in the two cases.
Hence one cannot treat the effect of $\delta$ perturbatively.

There are certain difficulties in making numerical prediction at
the one loop level. As inputs we need the chiral limit values of
the parameters $D$, $F$, ${\cal C}$ and ${\cal H}$.  Consistency
requires that these values be extracted from experimental data
by using chiral perturbation theory results calculated at the
one loop level. As we have discussed earlier, one loop
calculations inevitably lead to requiring new and undetermined
terms of chiral power $2$.

There are serious ambiguities of a different nature involving
the coefficients $\cal{C}$ and $\cal{H}$.  The latter can be
determined only from the experimental value of the $\pi \Delta
\Delta$ vertex, etc. Needless to say,
no such data exists.  Hence one must either rely on models or
use the results of chiral perturbation theory itself to fix
$\cal{H}$.  One such approach has been pursued in Ref.
\cite{ButSavSp}, although without having included neccessary
counter--terms.
The quantity $\cal{C}$ can be determined from the decay width of
the decuplets. In principle, we should use the value of
$\cal{C}$ in the chiral limit.  Consistency requires that the
decay width be calculated in chiral perturbation theory at the
one loop level and compared with the experimental value to
extract its chiral limit.

We follow the strategy\cite{JenkMan2} of extracting ${\cal C}$
using the full (unapproximated) phase--space expression for the
decay width
\begin{equation}
\Gamma  =
\frac{{\cal C}^2 \lambda^{3/2} ((M_{10}^2 + M_{8}^2 - m^2 )}
{192 \pi f^2_\pi M_{10}^5}
\label{eval2}
\end{equation}
where $\lambda$ is the usual phase--space factor
\begin{equation}
\lambda = M_{10}^4 + M_{8}^4 + m^4  - 2 M_{8}^2 M_{10}^2 -
2 m^2 M^2_{8} -2 m^2 M^2_{10}.
\end{equation}
The average value obtained is ${\cal C}^2=2.56$\cite{JenkMan2}.

One should note an isssue which arises from the use of the heavy
baryon limit.  The latter gives the formula:
\begin{equation}
\Gamma  = 2 Im\{M_{10}\} =
\frac{{\cal C}^2 (\delta^2 - m^2 )^{3/2}}{12 \pi f^2_\pi}.
\label{eval1}
\end{equation}
For example, for the case of $\Delta\rightarrow N + \pi$, using
$\delta = 292\,MeV$ and $\Gamma=120.\,MeV$ , one obtains from
Eq. (\ref{eval1}) that ${\cal C}^2 = 1.2$, while from Eq.
(\ref{eval2}) that ${\cal C}^2 = 2.2$.  The difference between
these two evaluations, nearly a factor of two, arises from what
are formally $1/M_B$ corrections (Eq. (\ref{eval1}) is indeed
the $M_B \rightarrow \infty$ limit of (\ref{eval2})).  They are
nevertheless not small and would have to also be borne in mind
when going to higher power.

We present in Table~\ref{tablex} values of the mass combinations
which appear in GMO and Decuplet Equal Spacing (DES) rules.
Several sets of parameters have been used for the purpose of
comparison.
\newpage
\begin{table}[tbp]
\begin{center}
\begin{tabular}{|c|c|c|c|c|c|}
Quantites&Set 1&Set 2&Set 3&Set 4&Set 5 \\ \hline
$D$&$0.61$&$0.56$&$0.61$&$0.61$&$0.61$\\ \hline
$F$&$0.40$&$0.35$&$0.40$&$0.40$&$0.40$ \\ \hline ${\cal
C}^2$&$2.56$&$2.56$&$2.2$&$1.2$&$2.56$ \\ \hline ${\cal
H}^2$&$3.61$&$3.61$&$6.6$&$3.61$&$3.61$ \\ \hline $m_\pi
(MeV)$&$140.$&$140.$&$140.$&$140.$&$0.$ \\ \hline
$GMO(MeV)$&$10.0$&$8.8$&$9.0$&$6.0$&$12.2$ \\ \hline
$DES(MeV)$&-4.2&-4.2&27&14.7&-3.9 \\
\end{tabular}
\end{center}
\caption{The value of $\delta$ is $226\,MeV$ throughout.
The experimental value of $GMO$ is $6.5\,MeV$
and the average value of the violation of $DES$ is $27\,MeV$. }
\label{tablex}
\end{table}
The parameters of set 1 are those used by
Jenkins~\cite{Jenkins}.  The sets 2, 3 and 4 are designed to
show the dependence of the results on the parameters $D$, $F$,
${\cal C}$ and ${\cal H}$.  The set 5 shows the effect of
 zero pion mass, which comes almost entirely from the change
in the mass of $\eta$ as given by Eq.~(\ref{etamass}).  It is
clear that unlike GMO, the violations to the DES is
particularly sensitive to the parameters ${\cal C}$ and ${\cal
H}$.  Since ${\cal H}$ can only be experimentally extracted
through loop corrections, the importance of a consistent
one--loop evaluation of all the parameters entering the chiral
lagrangian must be emphasized.  Set 3 contains a typical set of
parameters which fit the average violation to the DES rule.

\section{Conclusions}

We have calculated the combinations of baryon masses,
given by Eqs.~(\ref{SU3GMO}) and (\ref{SU3DES}), which appear in
the GMO and DES rules, at one--loop-${\cal O}(p^3)$ level in  chiral
perturbation theory.  At this level, these combinations depend neither on any
counterterms nor on a renormalization scale $\mu$. Some ambiguities
remain concerning the values of the coupling constants. We find that one
cannot calculate the mass combinations at one--loop-${\cal O}(p^4)$ level
because of the presence of undetermined counter--terms required to handle
 ultraviolet divergences of two varieties.
One class of divergences arise from the chiral expansion and involves
both wavefunction and vertex renormalization.  The other class of
 divergences arise from the $1/M_B$ corrections of graphs
containing internal decuplet lines.  The $1/M_B$ corrections
also give rise to terms which are finite but non-analytic in $m^2$.
Unfortunately they include a dependence on interactions which arise only when a
decuplet is off its mass shell.  The associated coupling constants are
not known at present.

\bigskip
{\centerline{ACKNOWLEDGEMENTS}} Many thanks to Dave Griegel for
his discussions concerning the decuplet.  We also thank Ulf-G.
Mei{\ss}ner for bringing to our attention the work of Bernard
{\it et al.}, Ref. \cite{BerMeis1}.  This work was supported in
part by DOE Grant DOE-FG02-93ER-40762.
\newpage
\section{Appendix A}
It might be illuminating, especially for the issue of $1/M_B$
corrections, to describe one method of evaluation of the
integral in Eq. (\ref{mshift1}) using dimensional
regularization.  An alternative approach, with of course the
same result, can be found in \cite{Jenkins}.  Using standard
replacements for the nucleon propagator in terms of real and
imaginary parts, Eq. (\ref{mshift1}) can be rewritten as
\begin{eqnarray}
\delta M_B\left|_{M_B \rightarrow \infty} \right. &=& \frac{-i \beta}{2
f_\pi^2}
\int \frac{d^4 k}{(2 \pi)^4} \frac{k^2}
{(k^2 - m^2  + i\eta)} \times\nonumber\\ &&\left(P\frac{1}{v
\cdot k} - i \pi \delta(v
\cdot k)\right).
\label{mshift2}
\end{eqnarray}
Note that the integrand arising from the principle valued part
of the nucleon's propagator is odd under the transformation $k
\rightarrow -k$ and hence integrates to
zero.  Working in the nucleon's rest frame, the $k_0$ integral
is used to integrate over the delta--function.  Dimensional
regularization is then used for the remaining integrals over
space--like momenta.  We thereby obtain that
\begin{eqnarray}
\delta M_B\left |_{M_B \rightarrow \infty}\right. &=& -\frac{\beta}
{4 f_\pi^2 (2 \pi)^3}
\int d^3 \vec{k} \frac{\vec{k}^2}{\vec{k}^2 + m^2 }\nonumber\\
&=& -\frac{\beta}{4 f_\pi^2 (2 \pi)^3} \int d^3 \vec{k}
\frac{(\vec{k}^2+m^2 ) - m^2 }{\vec{k}^2 + m^2 }\nonumber\\
&=& \frac{\beta m^2 }{4 f_\pi^2 (2 \pi)^3}\int \frac{d^3
\vec{k}}
{\vec{k}^2 +m^2 }\nonumber\\ &=& \frac{\beta m^2 }{32
\pi^3 f_\pi^2} \pi^{3/2}
\Gamma(-1/2) (m^2 )^{1/2}\nonumber\\
&=& -\frac{\beta m^3 }{16 \pi f_\pi^2}.\label{mshift3}
\end{eqnarray}
The fact that the $1/M_B$ corrections to this result (given in
Eq.  (\ref{fullmass})) are small might have been anticipated
when noting that the singularity at $v \cdot k = 0$ in
(\ref{mshift2}) is not pinched.

\section{Appendix B}

The factor $a_R^{N, \Delta}(\mu)$ is the residual
finite piece of the counterterm in ${\cal L}_2^{\pi N}$ used to
renormalize the infinity in Eq.  (\ref{uvterms}) proportional to
$\delta^3$.  Explicitly, these are:
\begin{eqnarray}
{\cal L}_2^{\pi N} &\ni& \frac{\delta^3}{16 \pi^2 f^2_\pi}
\left(\frac53 \kappa + a_R^N \right)
Tr \overline{B} B\nonumber \\ &&+\frac{\delta^3}{16 \pi^2
f^2_\pi}\left(\frac23 \kappa + a_R^\Delta
\right) \overline{\Delta} \Delta
\label{deltathreect}
\end{eqnarray}
where $\kappa$ is an ultraviolet divergent constant given by
\begin{equation}
\kappa = {\cal C}^2\left(- \frac{1}{\epsilon} + \gamma_E -
{\rm ln}(4\pi) \right).
\end{equation}
The counterterms from ${\cal L}_2^{\pi N}$ that renormalize the
infinities in Eq. (\ref{uvterms}) proportional to $\delta
m^2 $ are:
\begin{eqnarray}
{\cal L}_2^{\pi N} &\ni& \frac{3 \delta \zeta}{32 \pi^2 f_\pi^2}\left(
\frac{1}{3}\kappa^\prime +
b_D^\delta\right) Tr \overline{B} \{\xi^\dagger M \xi^\dagger +
\xi M
\xi, B\}\nonumber\\
&&+ \frac{3 \delta \zeta}{32 \pi^2 f_\pi^2}\left(
\frac{-5}{18}\kappa^\prime +
b_F^\delta\right) Tr \overline{B} [\xi^\dagger M \xi^\dagger +
\xi M
\xi, B]\nonumber\\
&&+ \frac{3 \delta \zeta}{32 \pi^2 f_\pi^2}\left(
\frac{-8}{9}\kappa^\prime +
\sigma^\delta\right) Tr M(\Sigma + \Sigma^\dagger) Tr
\overline{B} B\nonumber\\
&&+ \frac{3 \delta \zeta}{32 \pi^2 f_\pi^2} (\tilde{m}- m_s)
\left(\frac{-1}{27} \kappa^\prime \right)
Tr (\Sigma + \Sigma^\dagger) Tr \overline{B} B\nonumber\\ &&+
\frac{3 \delta \zeta}{32 \pi^2 f_\pi^2}\left( \frac{1}{6}\kappa^\prime +
c^\delta \right) \overline{\Delta}( \xi^\dagger M \xi^\dagger +
\xi M
\xi) \Delta\nonumber\\
&&- \frac{3 \delta \zeta}{32 \pi^2 f_\pi^2}\left(
\frac{-1}{6}\kappa^\prime  +
\tilde{\sigma}^\delta \right) Tr M(\Sigma + \Sigma^\dagger)
\overline{\Delta} \Delta
\label{deltapict}
\end{eqnarray}
where the ultraviolent divergent constant $\kappa^\prime$ is
given by
\begin{equation}
\kappa^\prime= {\cal C}^2 \left( \frac{1}{\epsilon} - \gamma_E +
{\rm ln}(4\pi) + 1 \right).
\end{equation}
$\zeta$ is the proportionality constant in the
GMOR\cite{GMOR} relation, $\zeta = m^2_K/(m_s + \tilde{m}) =
m^2_\pi/2\tilde{m}$.  The set $\{b_D^\delta, b_F^\delta, \cdots
\}$ which enter in the
definition of $\alpha_i^\delta$ in Eq. (\ref{massformula}), are
the residual finite pieces of these counterterms.  Note that the
counterterms given in Eqs. (\ref{deltathreect}) and
(\ref{deltapict}) are the only terms from ${\cal L}_2^{\pi,N}$
that contribute to the baryon masses.

\begin{figure}
\vglue 2in

\caption{Contributions to the masses appearing at second order in the chiral
counting.  The crosses represent an insertion from ${\cal
L}_1^{\pi, N}$.  Fig. (a) is the contribution from wavefunction
renormalization, Fig. (b) represents a vertex correction; each
introduces counterterms needing specification.}
\label{msloop}
\end{figure}
\begin{figure}
\vglue 2in

\caption{The one--loop self--energy corrections to the baryon in which
the intermediate baryon is part of the same multiplet.  Dots
represent the goldstone mesons; a straight line, the baryon
octet; and a double bar, the baryon decuplet.}
\end{figure}
\begin{figure}
\vglue 2in

\caption{The one--loop chiral corrections to the baryon in which the
intermediate baryon is not part of the same multiplet. Notation
same as in Fig. (1).}
\end{figure}
\vglue 9.5in
\begin{table}
\begin{center}
\begin{tabular}{|c|l|}
$\alpha_N$ & $-2 b_D m_s - 2 \sigma (2 \tilde{m} + m_s) - 2 b_F
(\tilde{m} - m_s)$
\\ \hline
$\alpha_\Sigma$ & $- 2 \tilde{m} b_D - 2 \sigma( 2 \tilde{m} +
m_s )$\\ \hline $\alpha_\Xi$ & $-2 b_D m_s - 2 \sigma (2
\tilde{m} + m_s) + 2 b_F (\tilde{m} - m_s)$
\\ \hline
$\alpha_\Lambda$ & $\frac{2}{3}( \tilde{m} - 4 m_s) b_D - 2
\sigma( 2 \tilde{m} + m_s )$\\ \hline\hline
$\alpha_\Delta$ & $2 \tilde{m} c - 2 (2 \tilde{m} +
m_s)\tilde{\sigma}$\\ \hline $\alpha_\Sigma^*$ & $(\frac{2}{3}c
- 2\tilde{\sigma}) (2 \tilde{m} + m_s)$ \\ \hline $\alpha_\Xi^*$
& $ \frac{2}{3}c (\tilde{m} + 2 m_s) - 2 (2 \tilde{m} +
m_s)\tilde{\sigma}$\\ \hline $\alpha_\Omega$ & $2 c m_s - 2 (2
\tilde{m} + m_s)\tilde{\sigma}$\\
\end{tabular}
\caption{ The contributions to the baryon masses arising from
terms in ${\cal L}_1^{\pi N}$.  $m_s$ is the mass of the strange
quark and $\tilde{m}$ is the average mass of the up and down
quarks.}\label{tablealp}
\end{center}
\end{table}
\widetext
\vspace{.25in}
\begin{table}
\begin{center}
$8\rightarrow8\otimes8$ 
\begin{tabular}{|c|c|c|c|}
$B$&$\sum_{B'}\mid\langle B'\pi\mid B\rangle\mid^2$&
$\sum_{B'}\mid\langle B'K\mid
B\rangle\mid^2$&$\sum_{B'}\mid\langle B'\eta\mid
B\rangle\mid^2$\\ \hline
$N$&$3/2(F+D)^2$&$3F^2-2FD+5/3D^2$&$3/2F^2-FD+1/6D^2$\\ \hline
$\Sigma$&$4F^2+2/3D^2$&$2F^2+2D^2$&$2/3D^2$\\ \hline
$\Lambda$&$2D^2$&$6F^2+2/3D^2$&$2/3D^2$\\ \hline
$\Xi$&$3/2(F-D)^2$&$3F^2+2FD+50/3D^2$&$3/2F^2+FD+1/6D^2$\\
\end{tabular}
\vspace{0.25in}
$8\rightarrow10\otimes8$ 
\begin{tabular}{|c|c|c|c|}
$B$&$\sum_{B\frac32}\mid\langle B\frac32\pi\mid B\rangle\mid^2$&
$\sum_{B\frac32}\mid\langle B\frac32K\mid
B\rangle\mid^2$&$\sum_{B\frac32}\mid\langle B\frac32\eta\mid
B\rangle\mid^2$ \\ \hline $N$&4/3 ${\cal C}^2$&1/3 ${\cal
C}^2$&\\ \hline $\Sigma$&2/9 ${\cal C}^2$&10/9 ${\cal C}^2$&3/9
${\cal C}^2$\\ \hline $\Lambda$&${\cal C}^2$&2/3 ${\cal C}^2$&\\
\hline
$\Xi$&3/9 ${\cal C}^2$&${\cal C}^2$&3/9 ${\cal C}^2$\\
\end{tabular}
\vspace{0.25in}
$10\rightarrow8\otimes8$ 
\begin{tabular}{|c|c|c|c|}
$B$&$\sum_{B}\mid\langle B\pi\mid B\frac32\rangle\mid^2$&
$\sum_{B}\mid\langle BK\mid
B\frac32\rangle\mid^2$&$\sum_{B}\mid\langle B\eta\mid
B\frac32\rangle\mid^2$ \\ \hline $\Delta$&1/3 ${\cal C}^2$&1/3
${\cal C}^2$&\\ \hline $\Sigma$&5/18 ${\cal C}^2$&4/18 ${\cal
C}^2$&3/18 ${\cal C}^2$\\ \hline $\Xi$&1/6 ${\cal C}^2$&1/3
${\cal C}^2$&1/6 ${\cal C}^2$\\ \hline $\Omega$&&2/3 ${\cal
C}^2$&\\
\end{tabular}
\vspace{0.25in}
$10\rightarrow10\otimes8$ 
\begin{tabular}{|c|c|c|c|}
$B$&$\sum_{B\frac32}\mid\langle B\frac32\pi\mid
B\frac32\rangle\mid^2$& $\sum_{B\frac32}\mid\langle
B\frac32K\mid B\frac32\rangle\mid^2$&$\sum_{B\frac32}\mid\langle
B\frac32\eta\mid B\frac32\rangle\mid^2$ \\ \hline $\Delta$&25/54
${\cal H}^2$&5/27 ${\cal H}^2$&5/54 ${\cal H}^2$\\ \hline
$\Sigma$&20/81 ${\cal H}^2$&40/81 ${\cal H}^2$& \\ \hline
$\Xi$&5/54 ${\cal H}^2$&30/54 ${\cal H}^2$&5/54 ${\cal H}^2$\\
\hline
$\Omega$&&20/54 ${\cal H}^2$&20/54 ${\cal H}^2$\\
\end{tabular}
\end{center}
\caption{The coefficents $\beta$ and $\beta^\prime$ for the one--loop
contributions to the baryon masses.}
\label{tablebet}
\end{table}

\begin{references}
\bibitem{Griegel}
D. K. Griegel, Phys. Rev. {\bf C43}, 904 (1991)
(Note that the free lagrangian is missing an overall minus sign; we
thank DKG for this observation).
\end{references}
\end{document}